\documentclass[aps,prl,twocolumn,showpacs]{revtex4}

\usepackage{amsmath}
\usepackage{amssymb}
\usepackage[dvips]{graphics}
\usepackage[dvips]{graphicx}

\newcommand{\numunit}[2]{%
  \ensuremath{#1\,#2}}

\begin{document}

\title{Probing decoherence through Fano resonances}
  \author{Andreas B\"arnthaler}
\affiliation{Institute for Theoretical Physics, Vienna University of
  Technology, A--1040 Vienna, Austria, EU}
\author{Stefan Rotter}
\thanks{Corresponding author, email: stefan.rotter@tuwien.ac.at}
\affiliation{Institute for Theoretical Physics, Vienna University of
  Technology, A--1040 Vienna, Austria, EU}
\author{Florian Libisch}
\affiliation{Institute for Theoretical Physics, Vienna University of
  Technology, A--1040 Vienna, Austria, EU}
\author{Joachim Burgd\"orfer}
\affiliation{Institute for Theoretical Physics, Vienna University of
  Technology, A--1040 Vienna, Austria, EU}
 \author{Stefan Gehler, Ulrich Kuhl, and Hans-J\"urgen St\"ockmann}
\affiliation{Fachbereich Physik, Philipps-Universit\"at Marburg,
  Renthof 5, D-35032 Marburg, Germany, EU}

\date{\today}
 
\begin{abstract}
We investigate the effect of decoherence on Fano resonances in wave
transmission through resonant scattering structures. We show that the
Fano asymmetry parameter $q$ follows, as a function of the strength of
decoherence, trajectories in the complex plane that reveal detailed
information on the underlying decoherence process. Dissipation and
unitary dephasing give rise to manifestly different trajectories. 
Our predictions are successfully tested against 
microwave experiments using metal cavities with different absorption 
coefficients and against previously published data on transport
through quantum dots. These
results open up new possibilities for studying the effect of
decoherence in a wide array of physical systems where
Fano resonances are present.
\end{abstract}

\maketitle

One of the central issues of current research in quantum mechanics
is {\it decoherence} \cite{zurek}, i.e., the loss of coherence 
induced in a system by the interaction with its environment. 
Studying the ubiquituous effects of decoherence is not only of
fundamental interest for the understanding of the
quantum--to--classical crossover, but is the key to the realization of
operating quantum information devices which rely on long 
coherence times \cite{380082,birdreview}. 
To this end, decohering processes need
to be controlled and suppressed. In practice, however,  enumeration
and identification of sources of decoherence is already a challenging
task on its own (see, e.g.,
\cite{PhysRevLett.93.077003,gurvitz}). This is, in part, due to
the fact that 
different decoherence channels are difficult to distinguish
from one another since their influence on the observables of interest
is often very similar.

Decoherence in quantum systems is, typically, described within the
framework of an open quantum system approach by a quantum master
equation of, e.g., the Lindblad form \cite{lindblad}. 
In this framework the reduced
density operator, $\rho$, of the open system with Hamiltonian $H_S$
interacting with the environment through Lindblad operators $L_j$
evolves as,
\begin{equation}
i\dot{\rho}=\left[H_S,\rho\right]+i\left[\sum_j
  L^{\phantom \dagger}_j\rho L^\dagger_j-\frac{1}{2}\left(L^\dagger_jL^{\phantom \dagger}_j\rho+\rho L^\dagger_jL^{\phantom \dagger}_j\right)\right]\,.\label{eq:lindblad}
\end{equation} 
The system-environment interactions allow for a decohering, yet unitary
evolution of the system within the Markov approximation. In the
special case that only the last term of the coupling in
Eq.~(\ref{eq:lindblad}) 
$(\sim L^\dagger_j L^{\phantom \dagger}_j\rho+\rho L^\dagger_j
L^{\phantom \dagger}_j)$ is
present, interaction with the environment
is purely dissipative. The counter-term $(\sim L^{\phantom
  \dagger}_j\rho L^\dagger_j)$
acts as source and preserves the unitarity of the evolution.
Characterizing the system-environment
interaction for a given physical realization is one of the major challenges of
decoherence theory. In this letter we show that Fano resonances,
specifically the asymmetry parameter $q$, allow to disentangle
different decoherence mechanisms present in resonant scattering devices
such as quantum
dots~\cite{gurvitz,goeres2000,kobayashi2003,PhysRevLett.93.106803,gong}
(for a review see \cite{fanoreview}). The $q$ parameter follows, as a
function of the decoherence strength, trajectories in the complex
plane that are specific to the underlying environmental coupling. In
particular, dissipation and decoherent dephasing can be distinguished
from each other. 

\begin{figure}[!t]
  \centering
  \includegraphics[draft=false,keepaspectratio=true,clip,%
                   width=82mm]{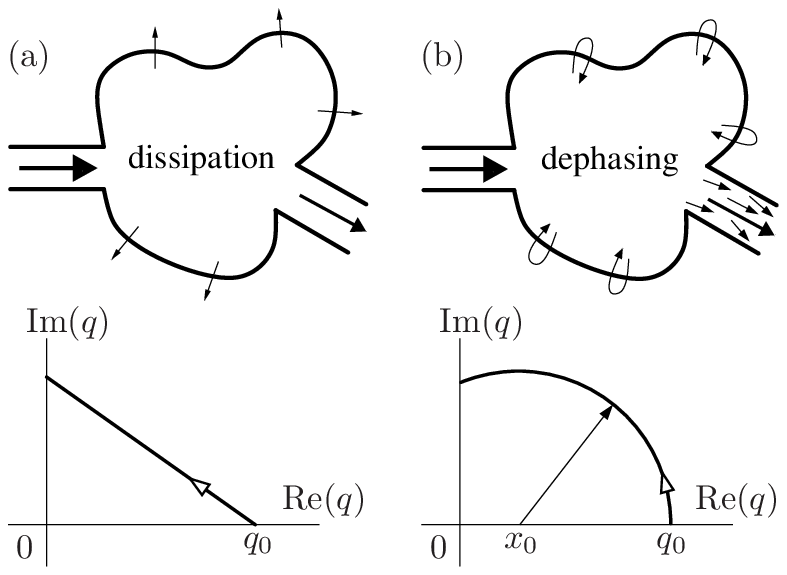}
  \caption{Dependence of the complex Fano asymmetry parameter 
    $q$ on the decoherence strength $\chi$. In the case of a uniformly
    dissipative system (a) an increase of $\chi$ shifts $q(\chi)$ in the
    complex plane along a \emph{straight line} away from a
    real value $q_0$ while for
    a uniformly dephasing system (b) the trajectory $q(\chi)$ forms a
    \emph{circular arc}.  This arc is centered at $x_0$ on the real
    axis thus featuring a vertical tangent at $q_0$.}
  \label{fig:q_ideal}
\end{figure}

Fano line shapes in transport result from the interference between
different channels in the transmission amplitude \cite{fano1961},
\begin{equation}
  t(k) = z_r\,\frac{\Gamma/2}{k-k_{\rm res}+i \Gamma/2} + t_d\,,
  \label{eqn:trans_res_bg}
\end{equation}
with $t_d$ the amplitude of a (smoothly varying) direct (or
background) channel and $z_r$ the strength, $k_{\rm res}$ the position,
and $\Gamma$ the width of the resonant channel. The transmission
probability in the vicinity of the resonance, $|t(k)|^2$ takes on the
form of a Fano profile,
\begin{equation}
  |t(k)|^2 = |t_d|^2 \frac{|\,\varepsilon+q\,|^2}{1+\varepsilon^2}, 
  \label{eqn:trans_prob}
\end{equation}
in terms of the reduced wavenumber (or energy) $\varepsilon=(k-k_{\rm
  res})/(\Gamma/2)$. The asymmetry parameter $q$ determines the shape
of the Fano resonance. In the limit $q\!\to\!\pm\infty$, the symmetric
Breit-Wigner shape is recovered while for $q\!\to\! 0$ window (or
``anti'') resonances appear. For single-channel scattering through
systems with time-reversal symmetry (TRS) $q$ is strictly real
\cite{footnote, lewenkopf}. When TRS is broken,
Eq.~(\ref{eqn:trans_prob}) still holds, but $q$ may take on complex
values \cite{clerk2001}. The generalization of the Fano $q$ parameter
is therefore ideally suited as a sensitive probe of TRS-breaking
processes. An Aharonov-Bohm ring exposed to a TRS-breaking
magnetic field was recently shown to exhibit $q$ parameters performing
periodic oscillations in the complex plane \cite{kobayashi2003}.
Decoherence, being a prime example for breaking TRS, should leave
distinct signatures in the behavior of $q$ as well
\cite{kobayashi2003,PhysRevLett.93.106803,clerk2001,rotter2005,zhang2006,gong}.
In the present letter we provide experimental and theoretical evidence
that the complex $q$-trajectory reveals details on the underlying
decoherence process that are characteristically different for
dissipation and irreversible dephasing.

We first consider ballistic transport 
through a scattering cavity in the presence of uniform dissipation,
the strength of which is independent of the wavelength or position
inside the cavity [Fig.~\ref{fig:q_ideal}(a)]. For the corresponding
open quantum system this corresponds to the reduction of the coupling
to the environment [Eq.~(\ref{eq:lindblad})] to sink terms $(\sim
L^\dagger_j L^{\phantom \dagger}_j\rho+\rho L^\dagger_j L^{\phantom
  \dagger}_j)$ only. Physical realizations of pure dissipation in
quantum dots include electron-hole recombination and currents leaking
into the substrate. For classical wave scattering \cite{jackson1998}
the presence of dissipation in a resonant device 
shifts the resonance positions
$k_{\rm res} \rightarrow \overline k_{\rm res}=$ $k_{\rm res} - \Delta
k-i\kappa$. Since, in relative terms, the
broadening of the resonance width $\kappa$ usually 
dominantes over $\Delta k$,
Eq.~(\ref{eqn:trans_res_bg}) is modified as follows,
\begin{equation}
  t(k) \approx z_r\,\frac{\Gamma/2}{k-k_{\rm res}+i \Gamma/2 + i\kappa} + t_d\,,
  \label{eqn:trans_res_bg_kappa}
\end{equation}
with a broadened resonance width ($\Gamma+2\kappa$). 

As convenient measure for the strength of decoherence 
we use the ratio 
$\chi=2\kappa/(\Gamma+2\kappa)$, with the limiting cases $\chi=0$ in
the absence of decoherence and $\chi=1$ for dissipation-dominated
broadening. With Eq.~(\ref{eqn:trans_res_bg_kappa}) the generalized
$q$ parameter in Eq.~(\ref{eqn:trans_prob}) now becomes,
\begin{equation}\label{eq:lin}
q(\chi)=q_0+\chi(i-q_0)\,,
\end{equation}
where $q_0$ is the real $q$ parameter in the absence of
dissipation. For increasing dissipation strength the complex Fano $q$
parameter follows a straight line trajectory in the complex plane [see
Fig.~\ref{fig:q_ideal}(a)] which, for large dissipation strength
($\chi \rightarrow 1$), approaches the limit $q=i$. The linear
form of $q(\chi)$ follows from the assumption entering
Eq.~(\ref{eqn:trans_res_bg_kappa}) that only the resonant but not
the direct amplitude ($t_d$) is affected by decoherence---an
assumption which generally holds well for resonant scattering devices
(including microwave cavities).

\begin{figure}[!t]
  \centering
  \includegraphics[draft=false,keepaspectratio=true,clip,%
                   width=\columnwidth]{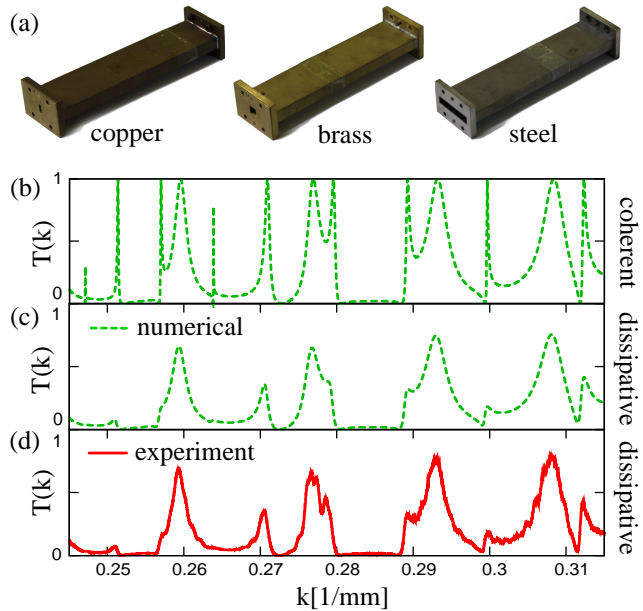}
  \caption{(Color online) (a) Rectangular microwave cavities of
    identical dimensions but 
    out of different materials. The left two cavities are shown with
    a shutter at the opening as used for tuning the Fano parameter.
    (b)-(d) Wavenumber dependence of transmission
    $T(k)$ through a cavity. In (b) the fully coherent limit is
    shown. (c) and (d) display the transmission with the dissipative
    decoherence as present in the steel cavity at room
    temperature. Theoretical MRGM (c) and experimental data (d) show
    excellent agreement on an absolute scale.}
  \label{fig2}
\end{figure}

The analytical dependence of $q$ on the decoherence strength has
previously been studied in the case of dephasing
[Fig.~\ref{fig:q_ideal}(b)] as the main source of
decoherence~\cite{clerk2001}. Here, in contrast to the dissipative
case, the flux in the system is conserved even at finite coherence
lengths. A simple realization of flux-conserving decoherent dephasing
for ballistic transport is given by the  B\"uttiker dephasing probe 
\cite{buettiker1988}: By attaching a fictitious
voltage probe, the coherent scattering paths
from source to drain are accompanied by incoherent paths
via the voltage probe which randomize the phase information.
To convert the dissipative voltage probe into a flux-conserving
dephasing probe, the potential of the probe is chosen such 
that the flux leaving through the probe is incoherently injected back into
the cavity. This corresponds to the presence of both sink and source
terms in the Liouvillian operator [Eq.~(\ref{eq:lindblad})] with an
infinite number ($j=1,\ldots,\infty$) of coupling terms with random
phases. Such an incoherent reinjection of flux is fully accounted for by an
additional Breit-Wigner shaped term in the Fano profile,
\begin{equation}\label{eq:6}
|t(k)|^2=|t_d|^2\left[\frac{\left\{\,\varepsilon'+{\rm Re}[q(\chi)]\,\right\}^2}{1+\varepsilon'^2}+\frac{{\rm Im}[q(\chi)]^2}{1+\varepsilon'^2}\right]\,,
\end{equation}
with the reduced wavenumber now rescaled to the increased resonance width $\varepsilon'=(k-k_{\rm res})/(\Gamma/2+\kappa)$ as well as ${\rm Re}[q(\chi)]=q_0(1-\chi)$ 
and ${\rm Im}[q(\chi)]^2=\chi\left[1+q_0^2(1-\right.$ $\left.\chi)\right]$.
By eliminating $\chi$ from these expressions, one finds
\begin{equation}\label{eq:qxqy}
\left\{{\rm Re}[q(\chi)]- x_0\right\}^2 + {\rm Im}[q(\chi)]^2 = r^2_0\,, 
\end{equation}
where $x_0 =[q_0 - 1/q_0]/2$ and $r^2_0 = 1 + x_0^2$ are independent
of $\chi$.  Thus $q(\chi)$ describes a circle in the complex plane
centered at $(x_0,0)$ on the real axis and converging to
$q(\chi\!\to\!  1)=i$ [see Fig.~\ref{fig:q_ideal}(b)]. 
We find circular trajectories in the complex $q$-plane also for a different 
(more general) scenario of dephasing modeled by gradually suppressing the
interference term between the direct and the resonant transmission  
in Eq.~(\ref{eqn:trans_res_bg}) as $\chi\!\to\!1$ (for details see appendix). 
Also here the circle is centered on the real axis, but imaginary values 
for $q(\chi\!\to\!1)$ may differ from $i$.

We conclude that for the same Fano resonance as
determined by the $q(\chi\!\to\! 0)=q_0$ limit, the complex generalization
of $q$ evolves along different trajectories for finite $\chi$ for
purely dissipative (on a straight line) and dephasing (on
a circular arc) decoherence. Even for small $\chi$ where $q$ is close
to the real axis
characteristic differences appear: the dephasing
trajectory has a tangent parallel to the imaginary axis while the
dissipative trajectory takes off at an angle $\arctan(1/q_0)$
relative to the $x$-axis. 
This finding suggests that by following
the $q(\chi)$ trajectory for a given Fano resonance, the underlying
decoherence process can be unambiguously identified. 

An ideal system
for the controlled experimental verification of the above
theoretical results are microwave cavities which have been
successfully employed in the past as analog simulators of a wide
variety of quantum transport phenomena \cite{stoeckmannbook}.
As was shown recently, well-separated Fano resonances can be
measured with high accuracy 
in such systems~\cite{rotter2005}.
The transport of microwaves into and out of the cavity can be
controlled via shutters at both ends, which in turn
determine the $q_0$ values of resonances. Due to the
finite conductivity of the cavity  and the resulting dissipation
of flux in the cavity walls, decoherence is
naturally present. Furthermore, we can
control the degree of dissipation by cooling the cavities to lower
temperatures or by fabricating cavities with identical geometry 
out of different materials.
To a good degree of approximation, the power loss can be assumed
to be uniform and mode-independent 
[as in Eq.~(\ref{eqn:trans_res_bg_kappa})].

For the experiment we used rectangular microwave
cavities (length $L=\numunit{176}${mm}, width
$D=\numunit{39}${mm}) [see Fig.~\ref{fig2}(a)]
made out of copper, brass and steel 
with different conductivities $\sigma$
[at room temperature:
$\sigma=\numunit{54.22}{\mathrm{m}/(\Omega\,\mathrm{mm}^2)}$ for copper,
  $\sigma=\numunit{12.20}{\mathrm{m}/(\Omega\,\mathrm{mm}^2)}$ for
  brass, and
  $\sigma=\numunit{1.37}{\mathrm{m}/(\Omega\,\mathrm{mm}^2)}$ for steel].
The cavities were terminated by two
metallic shutters each with opening width $s=\numunit{8.8}${mm} and a
thickness of $\numunit{1}${mm}. Two aluminum leads with length
$l=\numunit{200}${mm} and width $d=\numunit{15.8}${mm} were attached
to the openings. The measurements were carried out using a microwave
vector analyzer connected via coaxial cables and adapters at the end
of each lead. We note that the measuring device is thus ``part of the
semi-infinite lead'' and not an additional measurement channel 
\cite{gurvitz}. Measurements were performed in the frequency
range with exactly one transverse mode in each lead thus coupling to
the first and third transverse modes of the cavity (see
Ref.~\cite{rotter2005} for details). 

\begin{figure}[!t]
  \centering
  \includegraphics[draft=false,keepaspectratio=true,clip,%
                   width=\columnwidth]{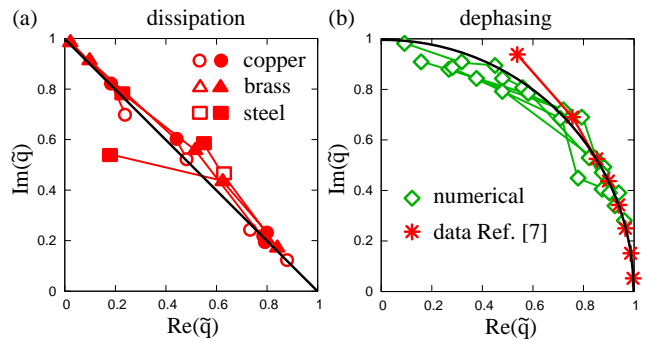}
                   \caption{(Color online) (a) Complex Fano $q$
                     parameter extracted from microwave experiments on
                     copper, brass and steel cavities.
                     For each material, measurements were performed at
                     room temperature (filled symbols) and liquid
                     nitrogen temperature (open symbols).  (b) Fano
                     parameter corresponding to the theoretical MRGM data for
                     a dephasing cavity (diamond symbols) and the
                     experimental data from Ref.~\cite{goeres2000} (asterisk
                     symbols). Data points for the same resonance at
                     different decoherence strengths are connected by
                     colored lines and rescaled, $q\!\to\!\tilde{q}$,
                     to reach $\tilde{q}=1$ $(\tilde{q}=i)$ in the
                     fully coherent (incoherent) limit.  The straight
                     [circular] black curve in (a) [(b)] displays the
                     theoretical prediction following
                     Eq.~(\ref{eq:lin}) [Eq.~(\ref{eq:qxqy})] for the
                     case of uniform dissipation [dephasing].  }
  \label{fig:q_rescaled_roomtemp}
\end{figure}

The experimental transmission data $T(k)=|t(k)|^2$ recorded for the
steel cavity display well-separated Fano resonances [see
Fig.~\ref{fig2}(d)].  For comparison we also show the corresponding
{\it numerical} results including dissipation [see Fig.~\ref{fig2}(c)]
and without dissipation [see Fig.~\ref{fig2}(b)]. The numerical data
was obtained with the modular recursive Green's function method
(MRGM)~\cite{rotter2000}, where uniform microwave attenuation by
dissipation following \cite{jackson1998} was taken into account.  Even
though the effect of dissipation is quite sizeable, we find excellent
agreement between theory and experiment [see Figs.~\ref{fig2}(c),(d)].
The small oscillations in the experimental data not reproduced by the
numerical calculations can be attributed to standing waves induced by
the minimal reflection from the adapters (less than 1\%).  
The influence on the Fano resonances is,
however, negligible as compared to the dominant decoherence process,
i.e., the ohmic losses in the cavity walls.  Accordingly,
Eq.~(\ref{eq:lin}) predicts the $q$ parameters of Fano resonances to
display linear decoherence trajectories in the complex plane. To
verify this prediction, we now extract the complex Fano $q$ parameter
from resonances at different decoherence strengths as determined by
the different cavity materials and their temperature dependence (on
each cavity one measurement was performed at ambient temperature and
at liquid nitrogen cooling). Since the Fano resonance formulas
Eqs.~(\ref{eqn:trans_res_bg}),(\ref{eqn:trans_prob}) are strictly only
valid for a direct amplitude $t_d$ which is $k$-independent, we
exclude resonances from our analysis for which this requirement is not
satisfied.  For resonances satisfying this requirement we extract the $q$
parameters by selecting the minimum, maximum and one intermediate
value as fitting points in the experimental transmission probabilities
[as, e.g., in Fig.~\ref{fig2}(d)]. Following each resonance for
different degrees of dissipation thus allows us to obtain the desired
decoherence trajectories in the complex $q$ plane. To facilitate the
comparison of the behavior of different members of the ensemble of
resonances we rescale all data [$q\!\to\!\tilde{q}={\rm
  Re}(q)\,q^{-1}_0+i\,{\rm Im}(q)$] onto a single ``universal''
trajectory which connects the points $\tilde{q}=1$ and $\tilde{q}=i$.
The data shown in Fig.~\ref{fig:q_rescaled_roomtemp}(a) demonstrates
that the expected linear behavior is, indeed, observed.

Such a linear trajectory due to pure dissipation can now be contrasted
with the circular trajectory for flux-conserving 
dephasing. For open quantum systems this corresponds to the presence
of an infinite number of source $(\sim L^{\phantom \dagger}_j\rho
L^\dagger_j)$ and sink terms $(\sim L^\dagger_j L^{\phantom
  \dagger}_j\rho+\rho L^\dagger_j L^{\phantom \dagger}_j)$ in
Eq.~(\ref{eq:lindblad}) as induced, e.g., by wave number independent
electron-phonon coupling. In the classical electromagnetic cavity such
a flux-restoring incoherent source is difficult to realize. However,
in our simulation we can numerically reinject the dissipated power
into the rectangular cavity, leaving the scattering system otherwise
unchanged. The dissipated flux to be symmetrically reinjected is
determined as the difference between transmission plus reflection and
the unitary limit, $1-|t(k)|^2-|r(k)|^2$. 
The $q$ values extracted
from the numerically determined Fano profiles are also mapped onto a
``universal'' circular arc with radius 1 centered at $x_0=0$. The data
obtained for several Fano resonances closely follow the circular
trajectory [Fig.~\ref{fig:q_rescaled_roomtemp}(b)] confirming the
dependence of the $q(\chi)$ trajectory on the underlying decoherence
mechanism. 

To test our predictions also for a true quantum scattering
system, we reanalyzed 
published experimental data on 
Fano resonances in transport through resonant quantum dots
\cite{goeres2000}. 
For these conductance measurements in the temperature range 
100mK $\le T\le$ 800mK, 
we find the evolution of $q(T)$ in the complex
plane to be very well described by a circular arc (see
Fig.~\ref{fig:q_rescaled_roomtemp}b)---as predicted for
flux-preserving dephasing.
Details of this analysis as well as a comparison between the
experimental and the theoretical Fano resonance curves are provided in
the appendix.

In conclusion, we have demonstrated that Fano resonances may serve as
sensitive probes of decoherence in wave transport. We find that for
increasing dephasing or dissipation strength the Fano asymmetry
parameter $q$ evolves on circular arcs or on straight lines in the
complex plane. As confirmed by measurements on microwave cavities and
on quantum dots,
these characteristic signatures provide a means to determine not only
the degree but also the specific type of decoherence present in the
experiment. It is hoped that the present findings will stimulate
future experimental investigations of the influence of decoherence
on the Fano $q$-parameter in resonant quantum transport.\\

\begin{acknowledgments}
  We thank K.~Kobayashi for very helpful discussions and the Austrian FWF
  (P17359 and SFB016) as well as the German DFG (FOR760) for support.
\end{acknowledgments}
\vspace{1cm}
\begin{appendix}{{\bf \large Appendix with supplementary material}}
\section{Complex $q$-trajectories: the case of dephasing}
Following previous analysis \cite{clerk2001} based on the B\"uttiker
dephasing probe model, we show in the main text of our article
that for this specific model the
Fano $q$-parameter follows a circular trajectory in the complex
plane. It is now instructive to inquire whether this result can also
be found for a more general scenario of dephasing. One such generic
approach is the ensemble average over a random phase $\phi$ between the
resonant ($\propto z_r$) and the background amplitude $t_d$,
\begin{equation}\nonumber
 t = t_d + \frac{z_r}{i+\varepsilon}\,e^{i\phi}\,,
\end{equation}
where the reduced wavenumber $\varepsilon=(k-k_{\rm res})/(\Gamma/2)$,
see Eq.~(\ref{eqn:trans_prob}). 
With the random phase $\phi$ featuring a zero mean
and a standard deviation $\sigma$ the interference term in the
ensemble average of the total transmission $\langle|t|^2\rangle_\phi$
will be gradually suppressed for increasing $\sigma$. This behavior
can be conveniently described by a prefactor $(1-\chi)\in[0,1]$,
containing the dephasing strength $\chi(\sigma)$,
\begin{equation}\nonumber
\begin{split}
  \langle|t|^2\rangle_\phi &= |t_d|^2 + \frac{|z_r|^2}{1+\varepsilon^2} + (1-\chi) \,2 \mathrm{Re}\left(t_d^* \,\frac{z_r}{i+\varepsilon}\right) \\
    &= |t_d|^2\, \frac{|z_r/t_d|^2 + (1-\chi)\,2 \mathrm{Re}\left[\frac{z_r}{t_d}(-i+\varepsilon)\right]+(1+\varepsilon^2)}{1+\varepsilon^2}\,.
\end{split}
\end{equation}
The limit of complete dephasing ($\chi \rightarrow 1$) corresponds
to the incoherent addition of resonant and background contribution.
With the real value of the Fano parameter $q_0$ in the absence of decoherence given as $q_0 = i+z_r/t_d$ \cite{clerk2001},
we further obtain,
\begin{equation}\nonumber
\begin{split}
  \langle|t|^2\rangle_\phi 
    = |t_d|^2 \,\frac{\left[\varepsilon+(1-\chi) q_0\right]^2-(1-\chi)^2 q_0^2 + q_0^2 +2\chi}{1+\varepsilon^2}\,.
\end{split}
\end{equation}
Comparing this expression with the general form of a Fano resonance as
in Eq.~(\ref{eq:6}) reveals the evolution of the complex Fano parameter
$q(\chi)$ as a function of the decoherence strength $\chi$,
\begin{align}\nonumber
  \mathrm{Re}[q(\chi)] &= (1-\chi)\,q_0\,,\\
  \mathrm{Im}[q(\chi)]^2 &= 2\chi + q_0^2\,(2\chi-\chi^2)\,. \nonumber
  \label{eqn:q_decoh}
\end{align}
Eliminating $\chi$ from the above two equations finally yields,
\begin{equation}\nonumber
  \mathrm{Im}[q(\chi)]^2 + \{\mathrm{Re}[q(\chi)] + 1/q_0\}^2 = 2 + \frac{1}{q_0^2} + q_0^2\,,
\end{equation}
which describes a circle in the complex plane with radius
$r_0=\sqrt{2+1/q_0^2+q_0^2}$, centered at $x_0=-1/q_0$ on the real
axis. Note that in the limit of complete dephasing ($\chi\to 1$)
the above trajectory $q(\chi)$ converges to a value on the imaginary
axis which, in contrast to the result obtained with the B\"uttiker dephasing
probe model \cite{clerk2001}, is not necessarily given by $q=i$.\\

We thus arrive at the interesting conclusion that different models of
dephasing may yield a similar circular behavior of $q(\chi)$ where,
however, the radius of the circular arc and its corresponding end
point depend on the specific dephasing scenario.

\begin{figure}
\includegraphics[
                   width=0.9\columnwidth]{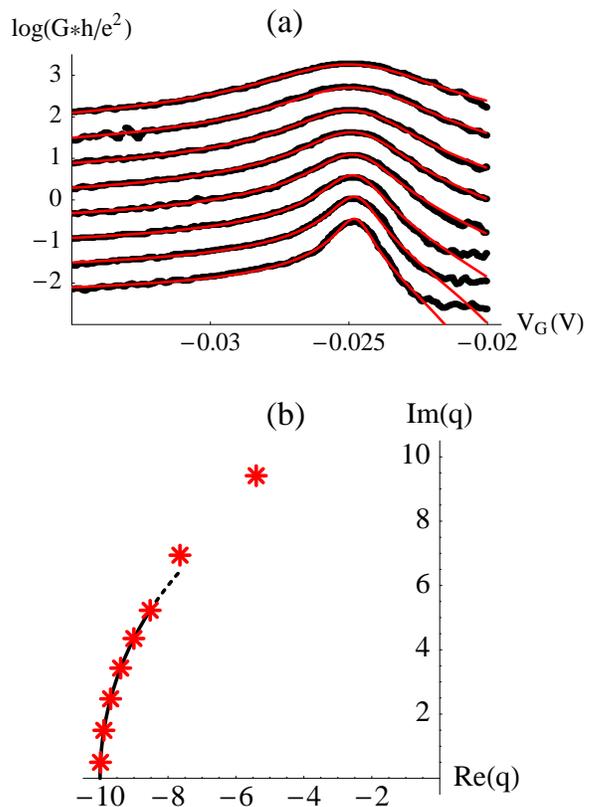}
                   \caption{(a) Voltage dependence of the conductance
                     through a resonant quantum dot as measured in the
                     experiment by Zacharia et
                     al.~\cite{goeres2000}. The black dots show the
                     values extracted from \cite{goeres2000} in steps
                     of 100 mK between 100mK (lowest curve) and 800mK
                     (highest curve). For better visibility the data
                     sets with $T>$ 100 mK are each multiplied by a
                     factor of $\sqrt{10}$ from one temperature value
                     to the next.  The red curves show the analytical
                     curves resulting from restricted parameter fits
                     (see text) with the corresponding complex
                     $q$-values being displayed in (b) by the red
                     asterisk symbols.  The black curve in (b) shows a
                     circular arc as prescribed for the dephasing
                     model of decoherence.  }
\label{fig:fits}
\end{figure}

\section{Details on the employed fitting procedures}

When extracting the complex value of $q$ from a resonance profile, the
imaginary part of $q$ is highly sensitive to the minimum value of the
profile. General automated fitting routines, however, do typically not
account for this specific dependence appropriately. To overcome this
difficulty we used the minimum, maximum, and an intermediate point of
the resonance as interpolation points in our procedure. Furthermore,
we took advantage of the knowledge that the minimum and maximum points
are extreme values of the Fano resonance, resulting in altogether five
equations.  Their solution yields the resonance amplitude, the
resonance position $k_{\rm res}$, the resonance width $\Gamma$ and the
real and imaginary part of $q$. All the $q$-values in
Fig.~\ref{fig:q_rescaled_roomtemp}
corresponding to the microwave experiment and to the numerical
dephasing data were extracted in this way. Resonances with a
non-uniform background were excluded from our analysis.

As outlined in the main text, we also tested our predictions against
experimental data previously published in \cite{goeres2000}. In that
publication the resonant conductance through a quantum dot was studied
as a function of temperature $T$ (100mK $\le T \le$ 800mK). The
corresponding experimental data extracted directly from
\cite{goeres2000} are shown in Fig.~\ref{fig:fits}(a) (black dots), right
above. Unfortunately, for all the resonance curves provided in
\cite{goeres2000} the background transmission is very small, such that
the corresponding Fano resonance lineshapes in the conductance are
very near the symmetric Breit-Wigner limit (with $|q|\gg 1$).  In this
limit we find rigorous multi-parameter fits with a complex-valued
$q$-parameter as performed on the more asymmetric Fano resonances
(with $|q|\lesssim 1$ as in the microwave data) to be unfeasible.
This is because for $|q|\gg 1$ the symmetry-restoring effect of
decoherence is hard to quantify. To test if the quantum dot data can
be described by our theoretical predictions, we instead performed a
consistency check whether a trajectory $q(\chi)$ can be found that
features good agreement with both the experimental data and a circular
form of $q(\chi)$.  For this purpose we carried out restricted
parameter fits which, in line with \cite{goeres2000}, were performed
on a logarithmic conductance scale with thermal broadening being
included separately and the data for the fit being restricted to
the range where the conductance is at least twice as large as the
measured conductance minimum (to reduce the influence of neighboring
peaks).
The best curves which we found in this way are shown in Fig.~\ref{fig:fits}(a)
(red lines) with the corresponding $q$-values shown in
Fig.~\ref{fig:fits}(b) (red symbols) and in
Fig.~\ref{fig:q_rescaled_roomtemp}(b) (after rescaling to the
unit circle). The overall very good agreement demonstrates that the
arc-like behavior of $q(\chi)$ can very well describe the experimental
data. To cross-check this result we also mapped the measurement data
on a linear $q(\chi)$-trajectory (not shown) as prescribed by the
dissipation-dominated decoherence in Eq.~(\ref{eq:lin}) and found much larger
discrepancies. We hope future quantum transport experiments will
make more asymmetric Fano resonances (with $|q|\lesssim 1$) available for
rigorous analysis.

\end{appendix}


\begin{thebibliography}{22}
\bibitem{footnote} Note that the $q$-parameter is complex for multi-channel
  scattering with TRS \cite{lewenkopf} (a case
  not considered here).
\end{thebibliography}
\end{document}